\documentclass[12pt,preprint]{aastex}
\usepackage{times}

\newcommand{\ms}{m~s$^{-1}$}

\shorttitle{Meridional Flow in Cycle 24}
\shortauthors{Zhao et al.}

\begin{document}
\title{Solar Meridional Flow in the Shallow Interior during the Rising 
Phase of Cycle 24}

\author{Junwei Zhao\altaffilmark{1}, A. G. Kosovichev\altaffilmark{2},
and R. S. Bogart\altaffilmark{1}}
\altaffiltext{1}{W.~W.~Hansen Experimental Physics Laboratory, Stanford
University, Stanford, CA~94305-4085}
\altaffiltext{2}{Big Bear Solar Observatory, New Jersey Institute of 
Technology, Big Bear City, CA~92314-9672}

\begin{abstract}
Solar subsurface zonal and meridional-flow profiles during the rising phase
of solar cycle 24 are studied using time-distance helioseismology technique.
The faster zonal bands in the torsional-oscillation pattern show strong
hemispheric asymmetries and temporal variations in both width and 
speed. The faster band in the northern hemisphere is located closer
to the equator than the band in the southern hemisphere, and migrates 
past the equator when the magnetic activity in the southern hemisphere  
is reaching maximum. The meridional-flow speed decreases substantially 
with the increase of magnetic activity, and the flow profile shows two 
zonal structures in each hemisphere. The residual meridional flow, after 
subtracting a mean meridional-flow profile, converges toward the activity 
belts and shows faster and slower bands like the torsional-oscillation 
pattern. More interestingly, the meridional-flow speed above latitude $30\degr$ 
shows an anti-correlation with the poleward-transporting magnetic flux, 
slower when the following-polarity flux is transported and faster when 
the leading-polarity flux is transported. It is expected that this 
phenomenon slows the process of magnetic cancellation and polarity reversal 
in the high-latitude areas.
\end{abstract}


\section{Introduction}
Solar interior rotation and meridional circulation are global-scale flows
closely related to solar dynamo and solar cycles \citep{cha10}, and have 
been widely studied using different helioseismic methods. Temporal 
variations of the rotation and meridional flow were a focus of study by 
many authors \citep[e.g.,][]{how00, vor02, cho01, bas03, hab02, zha04, 
kom05, gon08, kom13}. The global rotation exhibits faster and slower rotational
bands, known as torsional oscillation, with one lower branch migrating toward
the equator together with the activity belts and one higher branch 
drifting poleward in each hemisphere \citep{how00, how09}. For the temporal 
variation of the meridional flow, it was reported that the residual 
meridional flow, obtained after removal of a quiet-period profile, 
converged toward the solar activity belts in the shallow interior 
\citep[e.g.,][]{zha04, giz04}, but diverged from the activity belts 
in the deeper regions \citep[e.g.,][]{cho01, bec02}. Later, \citet{gon10} 
reported that the shallow converging pattern not just existed during 
the active period but also extended to the solar minimum years. Using 
non-helioseismic method by tracking motions of magnetic elements,
\citet{hat11} reported a similar converging pattern on the solar surface.
It was suggested that this phenomenon was related to the increased 
radiative loss in the activity belts \citep{giz08}. Moreover, it was 
found that the meridional flow speed was faster when the Sun was quiet 
and became slower when activities grew stronger \citep[e.g.,][]{bas03, 
zaa06}. \citet{hat10} also reported that the meridional-flow speed at 
the surface substantially increased after the activity maximum of cycle 
23, which they suggested might have caused the unusually long activity 
minimum after that cycle.

Solar cycle 24 so far shows unusually low activities and also exhibits
some unusual signs for the next cycle. Through comparing torsional oscillations
of cycles~23 and 24, \citet{how13} found that the internal solar rotation
in higher latitude slowed down in cycle 24 from its previous cycle. 
Using coronal \ion{Fe}{14} line observations, \citet{alt13} found 
that cycle 24 differed from previous cycles with no polar-crown prominences 
rushing to the poles. Meanwhile, observation of sunspots showed that the 
magnetic field strength of the sunspots was steadily decreasing since 
the beginning of the previous cycle \citep{liv12}. All these observations have 
demonstrated that cycle 24 is an unusual cycle, and the next cycle seems 
to be unusual, too. Therefore, it is important to study the temporal
evolution of the interior dynamics of the Sun in the hope that 
these properties can shed light on understanding this special solar cycle.

In this Letter, using the first 3.7 years data acquired by {\it SDO}/HMI 
during the rising phase of cycle 24, we study the torsional oscillation
and meridional-flow patterns from the surface to about 20~Mm in depth.
In addition to confirming some previously reported results, we find some new
properties of solar rotation and meridional flow. The faster band in the 
northern hemisphere migrates past the equator into the southern 
hemisphere by the time magnetic activities reach maximum in the southern
hemisphere. The meridional-flow speed above latitude $30\degr$ shows an 
anti-correlation with the poleward-transporting magnetic flux, slower
(faster) when the following-polarity (leading-polarity) flux is transported 
toward the pole. It is unclear whether these newly detected properties 
are common in most solar cycles or are particular just for this special 
cycle, but understanding the dynamic implications of these properties is
crucial to understand this solar cycle with particularly low activity level,
as well as the general problems like the magnetic polarity reversal and 
solar dynamo. We describe our data analysis procedure in \S2, present 
results in \S3, and discuss possible artifacts and implications of 
our results in \S4.

\section{Data Analysis}

The Helioseismic and Magnetic Imager \citep[HMI;][]{sch12a, sch12b} onboard 
the {\it Solar Dynamics Observatory} \citep[{\it SDO};][] {pes12} provides 
continuous observations of the full-disk photosphere. A time-distance 
data-analysis pipeline was developed and routine computations are carried 
out every day at the Joint Science Operations Center (JSOC) for {\it SDO} 
\citep{zha12a}. The pipeline provides subsurface flow field and acoustic 
wave-speed perturbation maps up to a depth of 20~Mm every 8 hours, covering 
most of the solar disk with an area of $120\degr \times 120\degr$. 
For details of the pipeline procedure, please see \citet{zha12a}. The 
analysis presented in this Letter, using the subsurface flow velocities 
directly from the time-distance pipeline, covers 2010 May 1 through 
2014 January 14, a total of 3.7 years during the rising phase of solar 
cycle 24. This period has a total of 50 Carrington rotations, from CR~2096 
(this rotation has 70$\%$ coverage, and all other rotations have full 
coverage) through CR~2145. 

For each Carrington rotation, we average subsurface flow fields for 
the east-west and north-south flow speeds at different depths. The 
average is performed along a $30\degr$-wide central meridian band 
so that the systematic center-to-limb effect in the meridional-flow 
measurements can be readily removed following the prescription by 
\citet{zha12b}. Once the north-south flow as a function of 
latitude is obtained, a few corrections are needed before these flows 
can truly represent meridional flow. First, the systematic center-to-limb 
effect is removed. The center-to-limb effect remains approximately unchanged 
through the analysis period, however, the correction should take into 
account of the annual B-angle variation. Second, the effect caused by 
$p$-angle, the angle between the solar rotational axis with the camera
axis, must be removed. The Venus transit on 2012 June 5 helped the HMI 
team to determine a $p$-angle of $0\fdg07$ (Jesper Schou, private 
communication), and the time-distance pipeline incorporated this 
correction starting on 2012 August 1. Results before that date were 
retrospectively corrected. The third systematic effect comes from 
the error in the Carrington elements, which also shows annual changes 
\citep{bec05}, and this effect is removed by subtracting a fit to an 
annual sinusoidal variation of the cross-equator flow speed. Subsurface 
meridional-flow profiles are obtained after all the above-mentioned 
effects are removed from the north-south flows. The annual B-angle 
variation also causes some effects, though small, in determining the 
rotation speed, and an annual sine fit is subtracted as well to obtain the 
subsurface rotation speed from the averaged east-west flow speed.

\begin{figure}
\epsscale{0.95}
\plotone{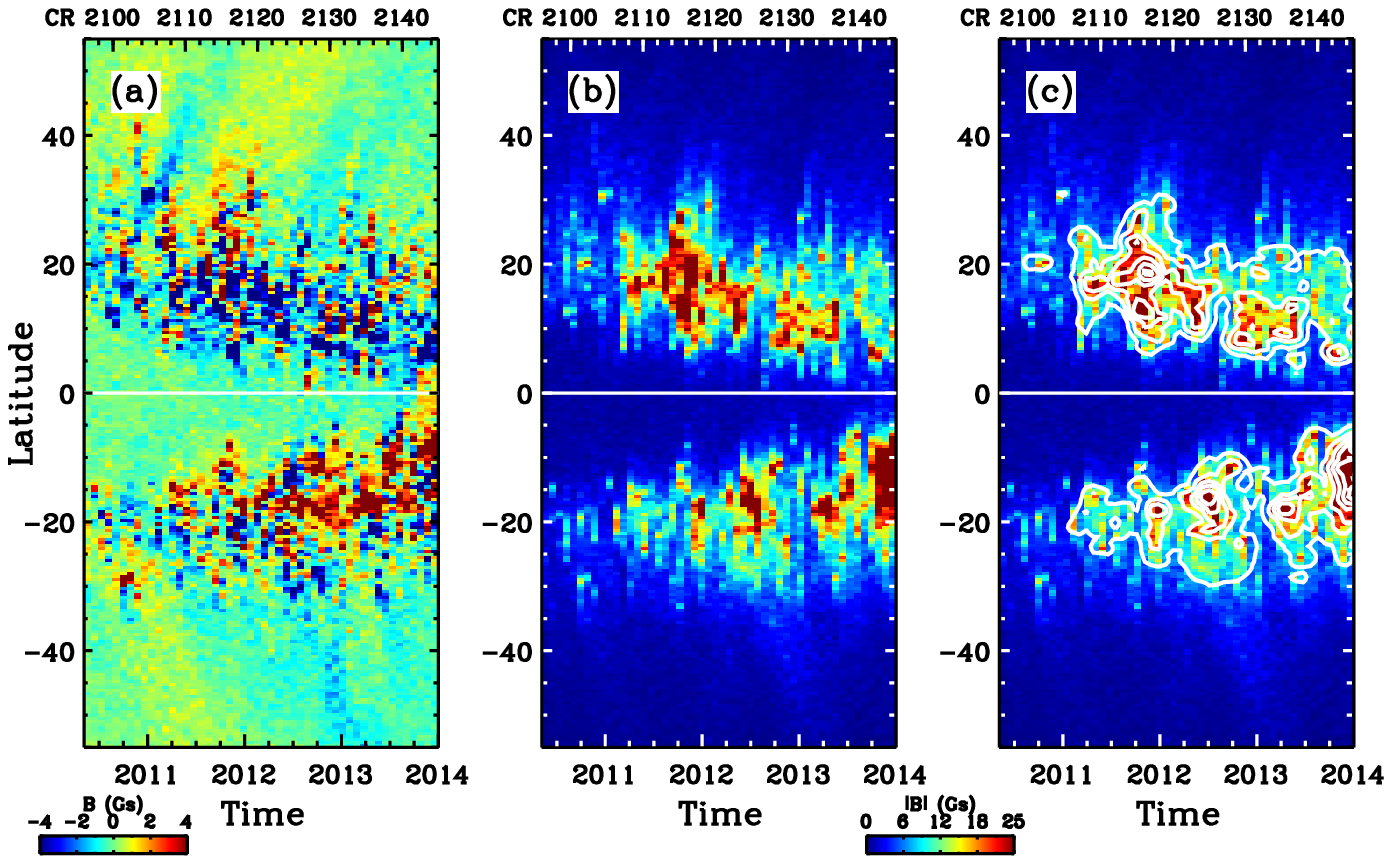}
\caption{(a) Butterfly diagram of net magnetic field, obtained by averaging 
over all longitudes for each Carrington rotation, for the analysis period.
(b) Same as panel (a) but for unsigned magnetic field. (c) Background image 
is the same as in (b), and contours show levels of 10, 15, 20, 25, and 30 
Gauss. For all panels here and in Figures~\ref{zonal} and \ref{merid}, 
the lower horizontal axis labels years and the upper horizontal axis labels 
Carrington rotation numbers.}
\label{mag}
\end{figure}

To better understand the connection of subsurface flow properties
with solar magnetic activity, the line-of-sight magnetic field from 
HMI is also averaged for each Carrington rotation to compare with the 
helioseismic results. Figure~\ref{mag} shows the longitudinally averaged
net magnetic field strength and unsigned magnetic field strength.
It is well known that for each 11-yr solar cycle, one hemisphere often 
has a leading magnetic polarity and a following polarity with opposite 
sign. The polarity switches sign in the cycle that follows. As shown 
in Figure~\ref{mag}a, for cycle 24 the leading polarity is negative 
(positive) in the northern (southern) hemisphere, and in each 
hemisphere more following-polarity magnetic flux is transported 
poleward than leading-polarity flux. Figure~\ref{mag}b shows that 
the northern hemisphere had its maximum activity during September through 
December 2011, and the activity in the southern hemisphere was reaching
its maximum during October 2013 through January 2014 (still ongoing 
when this analysis was done), nearly 2 years behind the maximum of 
the northern hemisphere.

\section{Results}

\subsection{Torsional Oscillation}
Torsional oscillation exhibits itself as zonal bands of faster and slower 
rotation, with the lower-latitude branches migrating toward the equator 
together with the activity belts, and the higher-latitude branches 
migrating toward the poles. Torsional oscillation is often obtained by 
subtracting a mean rotation profile from the rotation profile of each 
Carrington rotation. The mean profile is ideally obtained from a period 
of one solar cycle, and different mean profiles may result in different 
pictures of torsional oscillation \citep{how13}. In this study, as we 
only have 3.7 years data available, the mean rotation profile is averaged 
from the first 3 years, similar to what \citet{how13} did in their analysis. 
This may result in an imperfect determination of the torsional oscillation, 
but we limit our study to only the temporal variation. 

As shown in Figure~\ref{zonal}, the faster rotation band of the 
equatorward-migrating branch is on the equatorial side of the activity belts 
and the slower band on the poleward side. The torsional oscillation 
exhibits little change with depth, up to 21~Mm, consistent with results 
from the global-helioseismology analyses \citep[e.g.,][]{how13}. Our 
results also show that both the width and strength of the lower-latitude 
branch of the torsional oscillation change with time. The faster bands 
became weak and narrow after the activity maximum of the northern hemisphere, 
i.e., close to the end of the year 2011. The temporal variation of the
torsional-oscillation profile is qualitatively consistent with global 
helioseismology results using HMI data but not GONG data \citep[cf. both 
panels in Figure 9 of][]{kom14}.

Our torsional oscillation results also exhibits a hemispheric asymmetry. 
Throughout the analysis period, the faster band in the equatorward-migrating 
torsional-oscillation branch is on average approximately $3\degr$ closer 
to the equator in the northern hemisphere than in the southern hemisphere. 
Moreover, the faster band in the northern hemisphere extended past 
the equator into the southern hemisphere during the time the southern 
hemisphere was reaching its activity maximum. However, it is noteworthy 
that the recent ring-diagram analysis results did not seem to show 
this phenomenon \citep{kom14}.

\subsection{Decrease in Meridional-Flow Speed}

The meridional-flow profiles at selected depths averaged over one-year 
span for three consecutive years is shown in Figure~\ref{yearly}, and 
the profiles also exhibit a hemispheric asymmetry. The largest flow 
speed in the northern hemisphere, which showed strong magnetic activities 
2 yrs earlier than the southern hemisphere, is between $7\degr$ and 
$13\degr$, while the largest flow speed in the southern hemisphere is 
between $13\degr$ and $18\degr$, roughly $5\degr$ farther away from 
the equator. With the increase of magnetic activity, meridional-flow 
speed decreases substantially at all depths above the latitude $20\degr$ 
in both hemispheres. From June 2010 through May 2013, the flow speed 
at the depth of 0 -- 1~Mm drops an average of $6.3\pm1.1$~\ms\ between 
$13\degr$ and $30\degr$ in the northern hemisphere, but only drops 
$3.6\pm0.7$~\ms\ between $18\degr$ and $35\degr$ in the southern 
hemisphere. At the deeper depth of 7 -- 10~Mm, the flow speed decreases 
even more than at the shallower depths. At some latitudes, $25\degr - 
40\degr$N and $30\degr - 45\degr$S, the meridional flow even reverses 
directions, but the amount of reversal is similar to the size of error bars.

The meridional-flow profile shows two zonal structures in each hemisphere, 
and the size of these zonal structures changes with the evolution of 
the solar cycle. This is particularly clear in the deeper depth shown in 
Figure~\ref{yearly}b.  Select the green curve in Figure~\ref{yearly}b as 
an example: there is one zone between latitudes $0\degr$ and flow velocity
minimum at about $27\degr$N, and another zone above $27\degr$N. The lower 
latitude zone becomes smaller while the magnetic activity belts migrate 
toward the equator. The zonal structures look similar to what \citet{sch13} 
recently reported, but their results were for a larger depth.

\subsection{Temporal Variation of Meridional Flow}

\begin{figure}
\epsscale{0.95}
\plotone{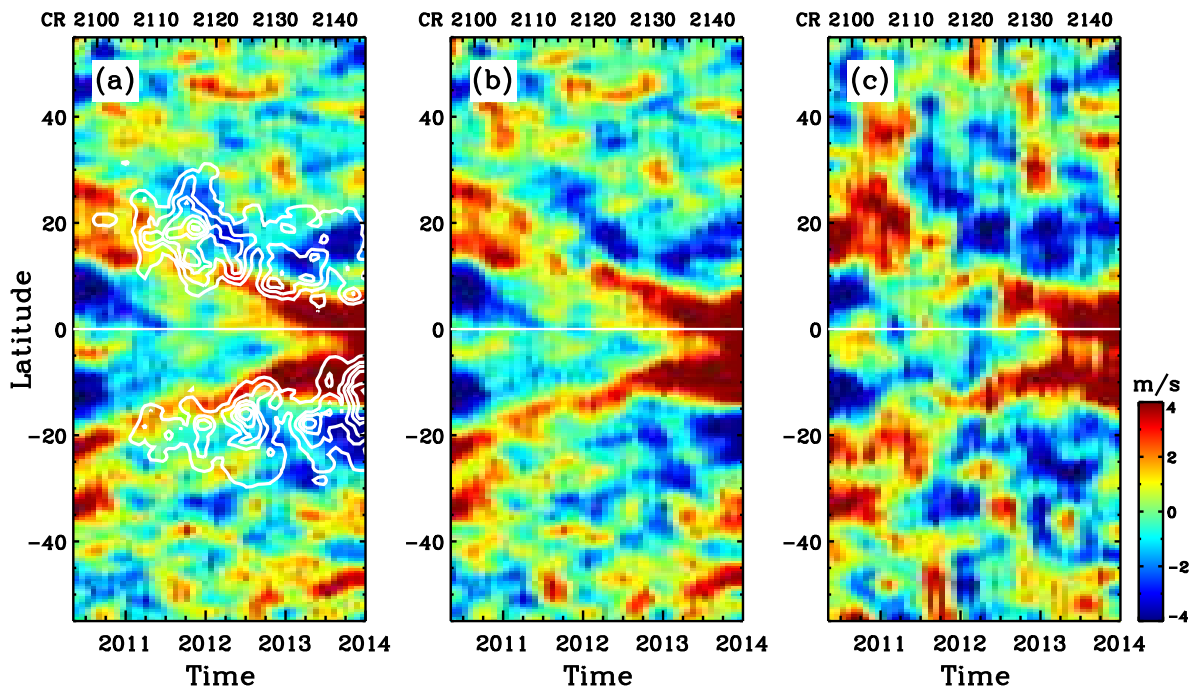}
\caption{Torsional oscillation showing zonal flow after subtracting the mean 
rotational flow profile, for the depths of (a) 0 -- 1 Mm, (b) 3 -- 5 Mm, and
(c) 17 -- 21 Mm. White contours in panel (a) are the same as those in 
Figure~\ref{mag}c, indicating the location of activity belts.}
\label{zonal}
\end{figure}

\begin{figure}
\epsscale{1.0}
\plotone{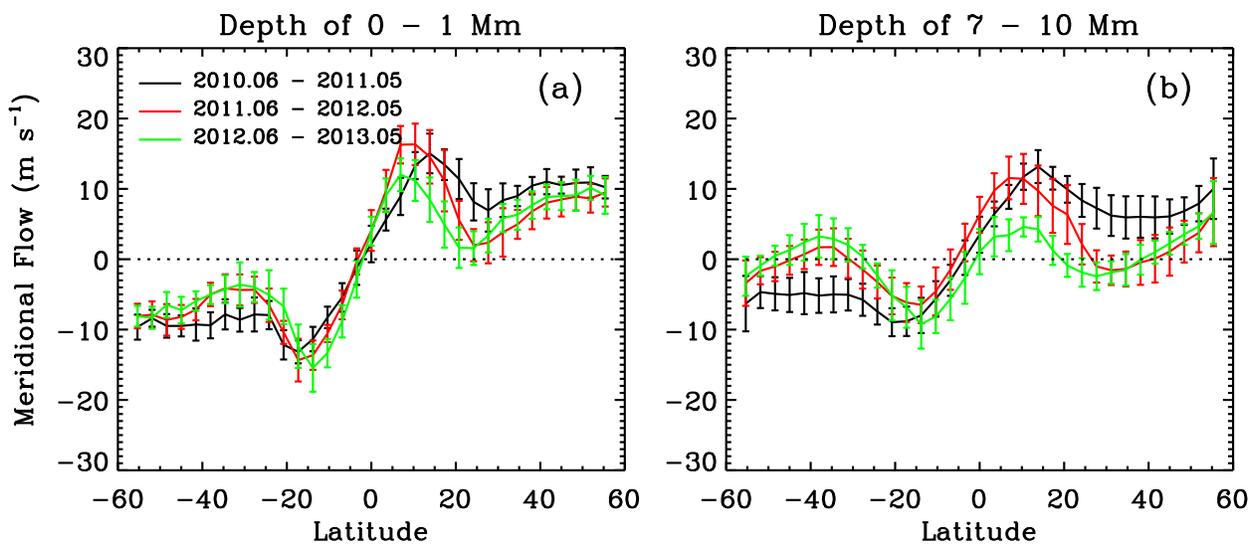}
\caption{Mean meridional-flow profiles averaged from three consecutive 
years (represented by different colors) for two selected depths (a) 0 -- 
1 Mm and (b) 7 -- 10 Mm. }
\label{yearly}
\end{figure}

\begin{figure}
\epsscale{0.95}
\plotone{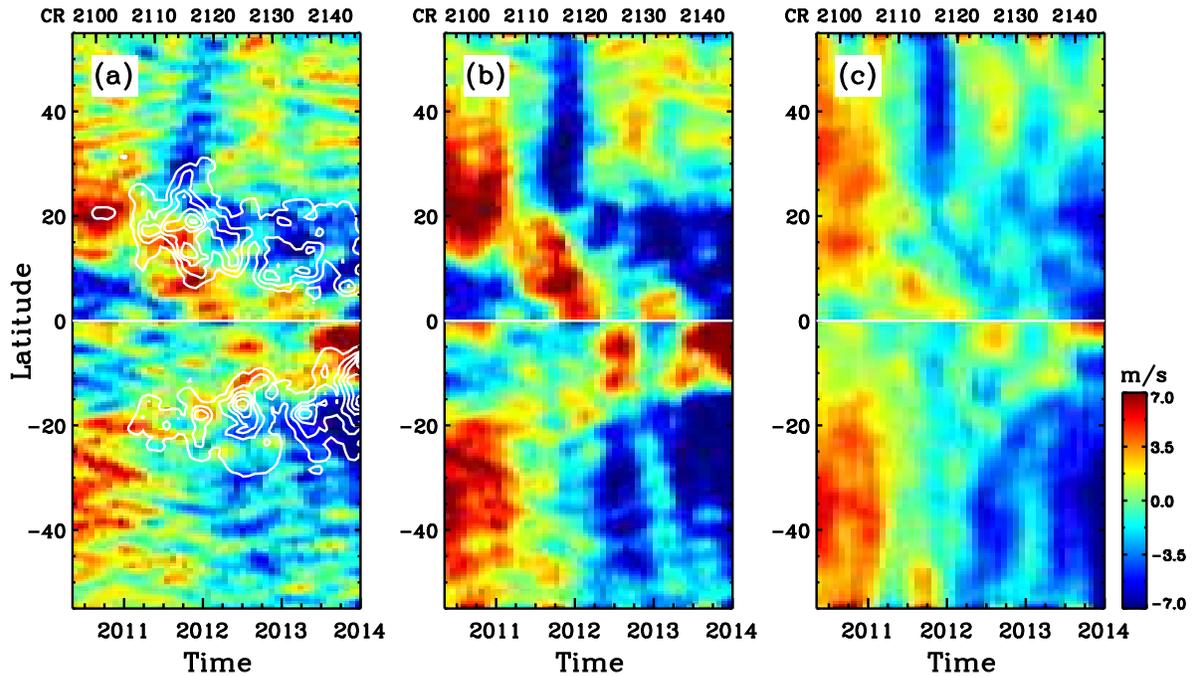}
\caption{Residual meridional flow after subtraction of the mean meridional-flow 
profiles, for the depths of (a) 0 -- 1 Mm, (b) 3 -- 5 Mm, and 
(c) 10 -- 13 Mm. The flow in the southern hemisphere is plotted with a 
reversed sign for a better visualization. Positive flow is poleward and 
negative flow is equatorward. White contours in panel (a) are the same 
as those in Figure~\ref{mag}c. }
\label{merid}
\end{figure}

Following the same procedure as for obtaining the torsional oscillation 
(Figure~\ref{zonal}), we subtract mean meridional-flow profiles, averaged 
from the first 3 years, from all the meridional-flow profiles of each 
Carrington rotation and get the residual meridional-flow profiles. 
Figure~\ref{merid} displays these residual flow profiles as functions of 
latitude and depth, with the residual flow in the southern hemisphere 
plotted with a reversed sign for a better visualization.

Similar to what was previously reported \citep[e.g.,][]{zha04, gon10}, 
the residual meridional flow converges toward the activity belts in both 
hemispheres. The residual flows also display faster (or poleward)
and slower (or equatorward) bands migrating toward the equator, just like
the torsional oscillation, up to a depth of 13~Mm. Between 13 and 21~Mm
our result does not show an organized pattern. Compared with the torsional 
oscillation (Figure~\ref{zonal}), the residual meridional-flow bands are 
more fragmented and wider, with the maximum speed $\sim$2~\ms\ faster than
the maximum torsional-oscillation speed. 

A comparison between Figure~\ref{mag}a and Figure~\ref{merid}a indicates 
that an anti-correlation exists between the residual-flow speed and the
net magnetic field strength above the latitude $\sim30\degr$ in
both hemispheres. For example, in the northern hemisphere, between CR~2110 and 
CR~2120, there is a branch of equatorward residual flow, and at about 
the same time there is a poleward transport of the following-polarity 
(positive) magnetic flux. A similar example can be found in the southern 
hemisphere between CR~2120 and CR~2130. These above-mentioned areas also show 
some enhancement in the unsigned magnetic field strength (Figure~\ref{mag}b), 
but the correlation between the residual-flow speed and the unsigned 
magnetic field, generally below 0.4 depending on latitude, is not as high 
as the correlation between the flow speed and the net magnetic field, 
up to 0.86 at the latitudes displayed in Figure~\ref{vy_mag}.

To better illustrate this anti-correlation between the meridional flow 
and the net magnetic field, we average both quantities along the latitudinal 
bands of $35\degr - 40\degr$ in the northern hemisphere and $40\degr 
- 45\degr$ in the southern hemisphere (see Figure~\ref{vy_mag}). The 
general trend for the meridional flow is that the speed decreases with 
the rise of magnetic activity, as already presented in \S3.2; however, 
superimposing on this general trend is a component of that the flow speed 
is anti-correlated with the net magnetic field. That is, the poleward flow 
speed is faster when the leading-polarity field is transported toward the 
pole and slower when the following-polarity flux is transported. It is 
true that these two quantities show better anti-correlation in the 
selected latitudinal bands than in the other bands above latitude $30\degr$, 
as the flow speed and the net magnetic flux become out of phase above 
the selected latitudinal bands (compare Figures~\ref{mag}a and \ref{merid}a).

\section{Discussion}
Through analyzing the subsurface flow fields obtained from the HMI 
time-distance data-analysis pipeline, we have studied the zonal-flow
and meridional-flow profiles during the rising phase of solar cycle 24.
The torsional oscillation from our analysis shows strong temporal variation
and hemispheric asymmetry, and the faster band in the northern hemisphere
is often closer to the equator than its southern counterpart, even 
extending past the equator near the end of the analysis period.
Meridional-flow speed above the latitude of $20\degr$ drops substantially 
with the rise of magnetic activity, and exhibits two zonal structures 
below latitude $55\degr$ in each hemisphere. From the surface to a depth 
of 13~Mm, the residual meridional flow, after a mean flow profile is 
removed, converges toward and migrates together with the activity belts. 
More interestingly, the meridional-flow speed above latitude $30\degr$ 
shows an anti-correlation with the net magnetic field, i.e., the poleward 
flow slows down (speeds up) when the following-polarity (leading-polarity) 
magnetic flux is transported poleward. 

We believe the results reported above are unlikely artifacts caused by
surface magnetic field due to the following reasons. Our results of torsional 
oscillation are qualitatively consistent with the results from global 
helioseismology, a method believed not contaminated by the surface magnetic 
field \citep{how99}. The converging residual meridional flow toward the 
activity belts is consistent with the non-helioseismic method of tracking 
motions of magnetic elements \citep{hat11}. Ring-diagram analysis presented 
evidence of that the converging flow trend extended beyond the active 
period into the quiet period \citep{gon10}, demonstrating that these 
properties were not due to the surface magnetism. In this study, the 
meridional-flow speed above $30\degr$, where no active regions appear, shows a
substantial anti-correlation with the net magnetic field, but shows little
correlation with the total unsigned magnetic field. Moreover, our experiments
with and without masking the regions, where magnetic field strength is
greater than 100~Gs, show that these magnetized regions do not cause a 
change of final results larger than the measurement uncertainties, mostly 
because the total area of these regions is just a small fraction of the Sun. 
All these evidences support that the above-reported results are truly 
solar phenomena instead of magnetic artifacts.

\begin{figure}
\epsscale{1.00}
\plotone{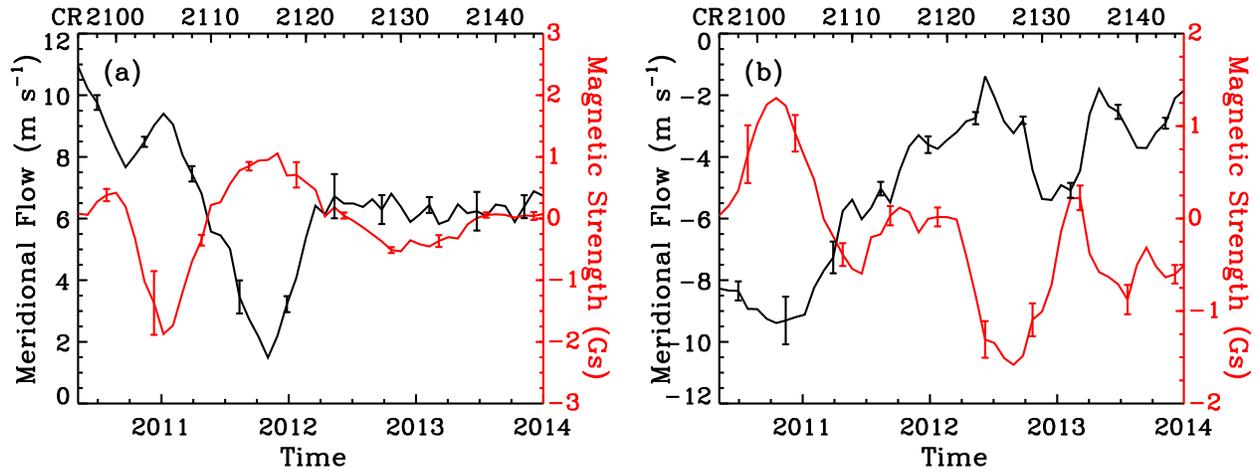}
\caption{(a) Black curves are meridional-flow velocity at the depth of 0 -- 1
Mm, as a function of time, averaged from the latitude $35\degr - 
40\degr$N. Red curves are magnetic field from the same latitudinal 
bands. (b) Same as in panel (a) but for the latitude $40\degr - 
45\degr$S. Note that the meridional flow speed increases with the vertical 
axis in panel (a) but decreases in panel (b), and the speed is displayed 
without the mean flow subtracted. }
\label{vy_mag}
\end{figure}

As shown in Figure~\ref{vy_mag}, the following-polarity magnetic flux, 
which is transported to the polar region and expected to cause the magnetic 
polarity reversal, is often associated with slower meridional-flow speed,
while the leading-polarity flux, which is transported poleward to strengthen
the high-latitude magnetic field, is often associated with faster flow 
speed. Since the helioseismically measured meridional-flow speed is 
similar to the flux transport speed \citep{sva07}, it is expected that 
this phenomenon slows down the magnetic cancellation in the high-latitude 
areas, hence delays the polarity reversal and the start of next solar 
cycle, according to the flux-transport dynamo theory \citep{cho95, dik99}. 
It is possible that the observed phenomenon of the poleward meridional-flow 
slow-down is more closely associated in time with the relatively strong 
magnetic activities in the lower latitude than associated with the 
poleward-transporting magnetic flux; however, the effective outcome 
is that the following-polarity magnetic flux is transported toward 
the pole with a lower speed than the leading-polarity magnetic flux. 
This is an interesting phenomenon worth further monitoring to see whether 
it continues in the late phase of cycle 24 and in other solar cycles. 
The physical cause of this phenomenon, how it affects the solar dynamo, 
and what it implies for the activity level of the next solar cycle are 
intriguing questions requiring more observational and modeling efforts.

We believe it is essential to examine past solar cycles to see whether
the phenomena reported in this Letter are common in most solar cycles
or particular for this cycle with unusually low activities. 
{\it SOHO}/MDI \citep[{\it Solar and Heliospheric Observatory} / Michelson 
Doppler Imager;][]{sch95} only had about two months continuous coverage each 
year during 1996 through 2010, making it difficult to examine these phenomena. 
{\it GONG} ({\it Global Oscillation Network Group}), which began
observations suitable for local-helioseismology analysis in 2002, 
seems to provide a unique opportunity to examine the previous solar cycle.

\acknowledgments
We thank the anonymous referee for thoroughly and carefully reading our
manuscript, as well as giving numerous constructive comments that help 
to improve the quality of this paper. SDO is a NASA mission, and HMI 
project is supported by NASA contract NAS5-02139 to Stanford University.

\newpage

\end{document}